\documentclass[oldversion]{adva}
\usepackage[round]{natbib}
\usepackage{deluxetable-mj}
\usepackage{longtable}
\usepackage{epsfig}
\usepackage[usenames,dvipsnames]{color}
\newcommand{\xnote}[1]{}
\newcommand{\ynote}[1]{{\protect\color{red}{\it #1}}}

\newcommand{\sonda}[1]{{\it {#1}\/}}
\newcommand{\swift} {\sonda{Swift}}

\newcommand{\simlt}{\,\hbox{\lower0.6ex\hbox{$\sim$}\llap{\raise0.2ex\hbox{$<$}}}\,}
\newcommand{\simgt}{\,\hbox{\lower0.6ex\hbox{$\sim$}\llap{\raise0.2ex\hbox{$>$}}}\,}

\begin{document}
\titlerunning{A~decade of GRB follow-up by BOOTES in Spain}%
\title{A~decade of GRB follow-up by BOOTES in Spain (2003-2013)}
%
\author{Martin Jel{\'{\i}}nek$^{1,2}$\thanks{email: \texttt{mates@iaa.es}}
Alberto~J.~Castro-Tirado$^{2,3}$\and  	
Ronan~Cunniffe$^2$\and			
Javier~Gorosabel$^{2,4,5,\dagger}$\and	
Stanislav~V\'{\i}tek$^6$\and		
Petr~Kub\'anek$^{7,8}$\and		
Antonio~de~Ugarte~Postigo$^2$\and 	
Sergey~Guziy$^2$\and 			
Juan~C.~Tello$^2$\and			
Petr~P\'ata$^6$\and			
Rub\'en~S\'anchez-Ram\'{\i}rez$^2$\and	
Samantha~Oates$^2$ \and			
Soomin~Jeong$^{9,2}$ \and		
Jan~\v{S}trobl$^1$ \and			
Sebasti\'{a}n~Castillo-Carri\'on$^{10}$\and	
Tom\'as~Mateo~Sanguino$^{11}$ \and		
Ovidio~Rabaza$^{12}$ \and			
Dolores~P\'erez-Ram\'{\i}rez$^{13,\dagger}$ \and	
Rafael Fern\'andez-Mu\~{n}oz$^{14}$ \and 
Benito~A.~de~la~Morena~Carretero$^{15}$ \and	
Ren\'e~Hudec$^{1,6}$ \and 		
V\'{\i}ctor~Reglero$^8$ \and			
Lola~Sabau-Graziati$^{16}$		
}
\authorrunning{Jel\'\i nek et al.}
\institute{
Astronomick\'y \'Ustav AV \v{C}R, Ond\v{r}ejov, (AS\'U AV \v{C}R), Ond\v{r}ejov, \v{C}R \and
Instituto de Astrof\'\i sica de Andaluc\'\i a, IAA-CSIC, 18008 Granada, Spain \and
Departamento de Ingenier\'{\i}a de Sistemas y Autom\'atica (Unidad Asociada al CSIC), Universidad de M\'alaga, 29010 M\'alaga, Spain \and
Unidad Asociada Grupo Ciencia Planetarias UPV/EHU-IAA/CSIC, Departamento de F\'{\i}sica Aplicada I, E.T.S. Ingenier\'{\i}a, Universidad del Pa\'{\i}s Vasco UPV/EHU, Alameda de Urquijo s/n, E-48013 Bilbao, Spain. \and
Ikerbasque, Basque Foundation for Science, Alameda de Urquijo 36-5, E-48008 Bilbao, Spain \and
\v{C}esk\'e Vysok\'e U\v{c}en\'{\i} Technick\'e, Fakulta Elektrotechnick\'a, (FEL \v{C}VUT), Praha, \v{C}R \and
Fyzik\'{a}ln\'{\i} \'{u}stav AV \v{C}R, Na Slovance 2, CZ-182 21 Praha 8, Czech Republic \and 
Image Processing Laboratory, Univ. de Valencia, Burjassot (Valencia), Spain \and
Institute for Science and Technology in Space, Natural Science Campus, SungKyunKwan University, 440-746, Suwon, Korea \and
Universidad de M\'alaga, Campus de Teatinos, M\'alaga, Spain \and
Departamento de Ingenier\'{\i}a de Sistemas y Autom\'atica, Universidad de Huelva, E.P.S. de La R\'abida (Huelva), Spain \and
Department of Civil Engineering, University of Granada, CP 18071, Spain. \and
Universidad de Ja\'{e}n, Campus las Lagunillas, 23071 Ja\'{e}n, Spain \and
Instituto de Hortofruticultura Subtropical y Mediterr\'anea "La Mayora" (IHSM-CSIC), Algarrobo (M\'alaga), 29750  Spain\and
Estaci\'on de Sondeos Atmosf\'ericos (ESAt) de El Arenosillo (CEDEA-INTA), Mazag\'on, Huelva, Spain \and
Divisi\'on de Ciencias del Espacio (INTA), Torrej\'on de Ardoz (Madrid), Spain 
}


\abstract{
This article covers ten years of GRB follow-ups by the Spanish BOOTES stations:
71 follow-ups providing 23 detections.  
Follow-ups by BOOTES-1B from 2005 to 2008 were given in the previous article,
and are here reviewed, updated, and include additional detection data points as
the former article merely stated their existence.  
The all-sky cameras CASSANDRA have not yet detected any GRB optical afterglows, but limits
are reported where available.
\\
\\
{\it Dedicated to the memory of Dolores P\'{e}rez-Ram\'{\i}rez and Javier Gorosabel, who passed away while this paper was in preparation}
}

\offprints{ \hbox{Martin Jel\'{\i}nek}} 

\date{Received  / Accepted }

\maketitle

\hyphenation{Blu-stin}

\section{Introduction}

Ever since the discovery of Gamma-ray bursts (GRB) in 1967
\citep{1973apj...182l..85k}, it was hoped to discover their counterparts at
other wavelengths. 
The early GRB-related transient searching methods varied (wide-field optical
systems as well as deep searches were being employed), but, given the coarse
gamma-ray-based GRB localizations provided, generally lacked either sensitivity
or good reaction time. 
%
The eventual discovery of GRB optical counterparts was done only when an X-ray
follow-up telescope was available on the Beppo-SAX satellite
\citep{1997natur.387..783c}. 
The optical afterglow could then be searched for with a large telescope in a
small errorbox provided by the discovery of the X-ray afterglow. 
The first optical afterglow of a Gamma-ray burst was discovered this way in
1997 \citep{1997natur.386..686v}.

Since then, astronomers have been trying to minimize the time delay between
receiving the position and the start of observations -- by both personal dedication
and by automating the telescope reaction. 
The ultimate step in automation, to minimize the time delay, is a full
robotization of the observatory to eliminate any human intervention in the
follow-up process.
This way, the reaction time can be minimized from $\sim$10\,minute limit that
can be achieved with a human operated telescope to below 10\,seconds.
With improvements in computational methods and in image processing speed, 
blind (non follow-up) wide-field methods are starting to be practical in
the search for optical transients.  Although limited in magnitude
range, they have already provided important observations of the optical emission
simultaneous to the gamma-ray production of a GRB \citep{2008Natur.455..183R}. 

Since 1997, the robotic telescope network BOOTES has been part of the effort to
follow-up gamma-ray burst events\citep{2004AN....325..679C}.
As of now, the network of robotic telescopes BOOTES consists of six telescopes
around the globe, dedicated primarily to GRB afterglow follow-up. 
We present the results of our GRB follow-up programme by two telescopes of the
network -- BOOTES-1B and BOOTES-2 and by the respective stationary very wide
field cameras (CASSANDRA). 
This text covers eleven years of GRB follow-ups: 71 follow-ups providing 21
detections.

Different instruments have been part of BOOTES during the years in question:
a~30\,cm telescope which was used for most of the time at BOOTES-1 station but
at periods also at BOOTES-2, the fast-moving 60\,cm telescope at BOOTES-2
(Telma), and also two all-sky cameras, CASSANDRA1 at BOOTES-1 and CASSANDRA2 at
BOOTES-2. Results from CASSANDRAs are included where available, without paying
attention to the complete sample.

This article is a~follow-up of a previous article: \cite{bible1} which provided
detailed description of evolution of BOOTES-1B, and analysis of efficiency of
a~system dedicated to GRB follow-up based on real data obtained during four
years between 2005 and 2008. This work is a~catalogue of BOOTES-1B and BOOTES-2
GRB observations between 2003 and 2013; it is complete in providing information
about successfully followed-up events, but does not provide analysis of missed
triggers as did the previous article. 

\subsection{BOOTES-1B
\label{bootes1b-hist} }


BOOTES-1 observatory is located at the atmospheric sounding station at El
Arenosillo, Huelva, Spain (at lat: 37$^\circ$06$'$14$''$N, long:
06$^\circ$44$'$02$''$W).
Over time, distinct system con\-fi\-gu\-ra\-ti\-ons were used, including also
two 8\,inch S-C telescopes, as described in \citet{bible1}, the primary
instrument of BOOTES-1B is a~D=30\,cm Schmidt-Cassegrain optical tube assembly
with a CCD camera. 
Prior to June 15, 2007, Bessel $VRI$ filters were being used as noted with the
observations, any observations obtained after this date have been obtained
without filter ($C$ or clear). 
We calibrate these observations against $R$-band, which, in in the case of no
color evolution of the optical counterpart, is expected to result in a~small
($\sim 0.1$\,mag) constant offset in magnitude.

\subsection{BOOTES-2 \label{bootes2-hist}}

BOOTES-2 is located at CSIC's experimental station La Mayora (Instituto de
Hortofruticultura Subtropical IHSM-CSIC) (at lat: 36$^\circ$45$'$33$''$N, long:
04$^\circ$02$'$ 27$''$W), 240\,km from BOOTES-1.
It was originally equipped with an identical 30\,cm Schmidt-Cassegrain
telescope to that at BOOTES-1B. 
In 2007 the telescope was upgraded to a~lightweight 60\,cm Ritchey-Chr\`{e}tien
telescope on a~fast-slewing NTM-500 mount, both provided by Astelco. 
The camera was upgraded at the same time to an Andor iXon $1024\times 1024$
EMCCD, and in 2012 the capabilities were extended yet again to low resolution
spectroscopy, by the installation of the imaging spectrograph COLORES of our
own design and construction \citep{colores}.
Bessel magnitudes are calibrated to Vega system, SDSS to AB.

\section{Optical follow-up of GRB events
\label{catalogue}
}

%
%
%
%

Here we will detail the individual results for each of the 23 events
followed-up and detected in 2003 -- 2013. Each GRB is given a~short
introductory paragraph as a~reminder of the basic observational properties of the
event. Although we do not discuss the properties at other wavelengths, we try
to include a~comprehensive reference of literature relevant to each burst. As
GCN reports usually summarise the relevant GCN circular traffic, we have
omitted the raw GCN circulars except for events for which a~GCN report or other
more exhaustive paper is unavailable.

Further 48 follow-ups which resulted in detection limits are included in tables
\ref{table-bible-b1} and \ref{table-bible-b2}, but are not given any further
attention. 

One by one, we show all the successful follow-ups that these telescopes have
performed during the first ten years of the \swift\/ era, and since the
transition of the BOOTES network to the RTS-2 \citep{rts2} observatory control
system, which was for the first time installed at BOOTES-2 in 2003, and during
the summer of 2004 at BOOTES-1.

%
%
%

\begin{figure}[b!]
	\begin{center} \label{lc050824} \includegraphics{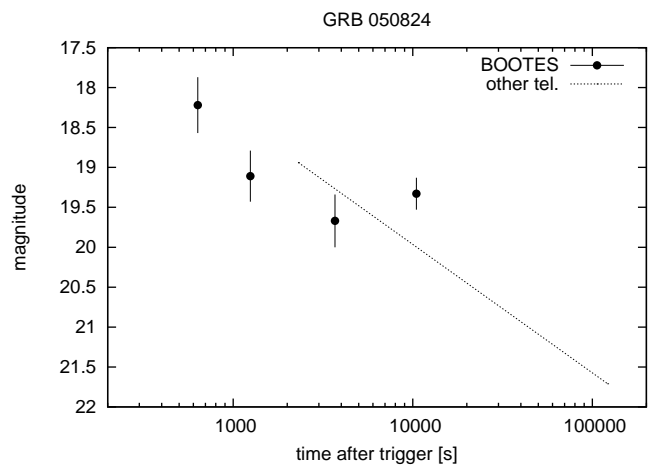}
	\caption{The optical light curve of GRB\,050824, the optical lightcurve represents behaviour seen by \cite{js050824}.}
	\end{center} 
\end{figure}

\paragraph{GRB\,050525A} 

A~bright low-redshift ($z=0.606$) localized by \swift\/ \citep{swift050525}.
Plenty of optical data, including the signature of the associated supernova
sn2005nc \citep{sn050525,resmi050525}.

GRB\,050525A was the first BOOTES-1B burst for which a~detection was obtained.
The telescope started the first exposure 28\,s after receiving the notice,
383\,s after the GRB trigger.  An optical afterglow with $V\simeq16$ was
detected. A~weak detection of a~bright GRB implied a~reexamination of observing
strategies employed by BOOTES. The largest, 30\,cm telescope was changed to
make R-band imaging instead of using the field spectrograph to greatly improve
sensitivity in terms of limiting magnitude. The 20\,cm telescopes were still
observing with V+I filters \cite[for details see][]{bible1}.  

This burst was covered in real time by both All Sky Cameras of BOOTES
(CASSANDRA1 and 2), providing an unfiltered limit of $>9.0$ \citep{aup050525}.
BOOTES observation of this GRB is included in \citet{resmi050525}.

\paragraph{GRB\,050824} 

A~dim burst detected by \swift. The optical afterglow of this GRB discovered
with the 1.5\,m telescope at OSN, redshift $z=0.83$ determined by VLT
\citep{js050824}.

BOOTES-1B was the first telescope to observe this optical transient, starting
636\,s after the trigger with $R\simeq17.5$. The weather was not stable and the
focus not perfect, but BOOTES-1B worked as expected.  In the end, several hours
of data were obtained. 
BOOTES observation of this GRB is included in \citet{js050824}.

\paragraph{GRB\,050922C}

 A~\swift\/ short and intense long burst \citep{swift050922c,bat050922c} that
was observed also by \sonda{HETE2} \citep{hete050922c}. Optical afterglow mag
$\sim 15$, $z=2.198$ \citep{z050922c}.

Due to clouds, the limiting magnitude of BOOTES-1B dropped from $\sim 17.0$ for
a~30\,s exposure to mere 12.9. The afterglow was eventually detected with the
$R$-band camera (at the 30\,cm telescope) during gaps between passing clouds.
The first weak detection was obtained 228\,s after the GRB trigger and gave
$R\simeq14.6$.

\begin{figure}[b!] 
	\begin{center} \label{lc051109}
	\includegraphics{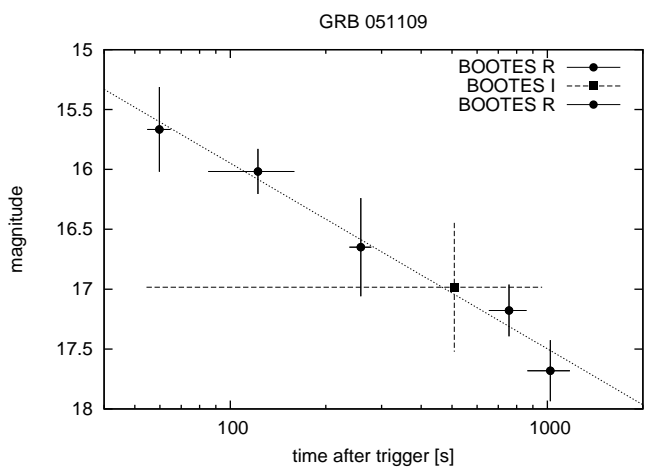}
	\caption{The optical light curve of GRB\,051109A. The dotted line
represents the optical decay observed by \citet{mirabal051109a}. }
	\end{center} 
\end{figure}

\paragraph{GRB\,051109A}

 A~burst detected by \swift\/ \citep{swift051109a}.  Optical afterglow mag
$15$, redshift $z=2.346$ \citep{z051109a}, optical lightcurve by
\citet{mirabal051109a}.

At BOOTES-1B the image acquisition started 54.8\,s after the burst with the
30\,cm telescope in $R$-band and one of the 20\,cm telescopes in $I$-band
\citep{mj051109a}. There were still a~number of performance problems -- most
importantly synchronization between cameras such that when the telescope
position was to be changed, both cameras had to be idle.  As the 30\,cm
telescope was taking shorter exposures, extra exposures could have been made
while waiting for the longer exposures being taken at the 20\,cm to finish. The
20\,cm detection is, after critical revision, only at the level of 2-$\sigma$.
The $R$-band observation shows the object until about 20 minutes after the GRB,
when it becomes too dim to measure in the vicinity of a~17.5\,m nearby star.
Mean decay rate observed by BOOTES is $\alpha=0.63\pm0.06$ ($F_{\mathrm opt}
\sim t^{-\alpha}$). 

The relatively shallow decay observed by BOOTES is in close agreement with what
was observed several minutes later by the 2.4\,m MDM ($\alpha=0.62\pm0.03$) and
according to an unofficial report \citep{web051109a} there was a decay change later, by about 3\,h
after the burst to $\alpha=0.89\pm0.05$. \looseness=-1

\begin{figure}[b!]
	\begin{center} \label{lc080330}
	\includegraphics{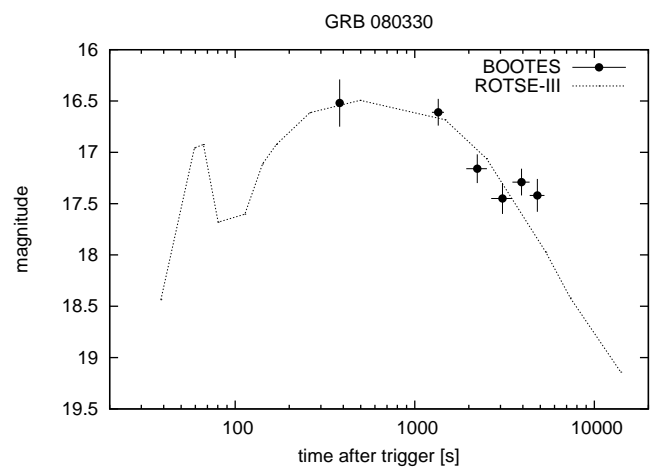}
	\caption{The optical light curve of GRB\,080330. The dotted line shows
how the lightcurve as seen by ROTSE-III \citep{rotse0330+0413a} }
	\end{center} 
\end{figure}

\paragraph{GRB\,080330}

 A~rather bright long burst detected by \swift\/ Afterglow reported to be
detected by UVOT, TAROT, ROTSE-III, Liverpool Telescope and GROND.
Spectroscopic redshift $z=1.51$ by NOT \citep{rep080330}.\looseness=1

This GRB happened during the first day recommissioning of BOOTES-1B after its
move from the BOOTES-2 site at La Mayora. The GCN client was not yet operational
and at the time of the GRB we were focusing the telescope. The first image was
obtained 379\,s after the GRB trigger and the optical afterglow was detected
with magnitude $\sim16.3$ on the first image. A~bug in the centering algorithm
caused a~loss of part subsequent data. Further detections were obtained
starting 21\,min after the GRB when the problem was fixed.  

The lightcurve (as seen by \citealt{rotse0330+0413a}) seems to show an
optical flare and then a possible hydrodynamic peak. The data of BOOTES,
however, trace only the final part of this behaviour, where the decay
accelerates after passing through the hydrodynamic peak.

\paragraph{GRB\,080413A} 

A~rather bright GRB detected by \swift, detected also by \sonda{Suzaku}-WAM,
optical afterglow by ROTSE-III \citep{rotse0330+0413a}, redshift $z=2.433$ by
VLT+UVES \citep{rep080413a}.

\begin{figure}[b!]
\begin{center} 
\includegraphics{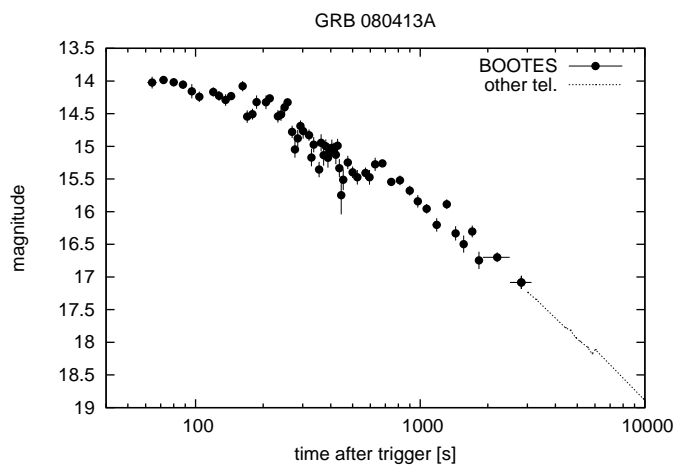}
\caption{The optical light curve of GRB\,080413A.
\label{bible-lc080413a} }
\end{center} 
\end{figure}

BOOTES-1B started obtaining images of the GRB\,080413A just 60.7\,s after the
trigger (46.3\,s after reception of the alert). An $R \simeq 13.3$ magnitude
decaying optical afterglow was found, see. Fig.\,\ref{bible-lc080413a}
(\citealp{pk080413a}, \ynote{Jelinek et al., in prep.})

\paragraph{GRB\,080430} 

A~burst detected by \swift. It was a~widely observed, low-redshift $z \simeq
0.75$ optical afterglow with a~slowly decaying optical afterglow
\citep{rep080430}. Observed also at very high energies by {\it MAGIC} without
detection \citep{magic080430}. 

BOOTES-1B obtained the first image of this GRB 34.4\,s after the trigger. An
optical transient was detected on combined unfiltered images with a~magnitude
$\simeq 15.5$ (\citealp{mj080430}). 
\xnote{No LC, no table?}\looseness=-1

\paragraph{GRB\,080603B} 

A~long GRB localized by \swift, detected also by \sonda{Konus}-Wind and by {\it
\sonda{INTEGRAL}} \citep{spiacs.cat}. Bright optical afterglow, extensive
follow-up, redshift $z=2.69$ \citep{rep080603b}. 

\begin{figure}[b!]
\begin{center} 
\includegraphics{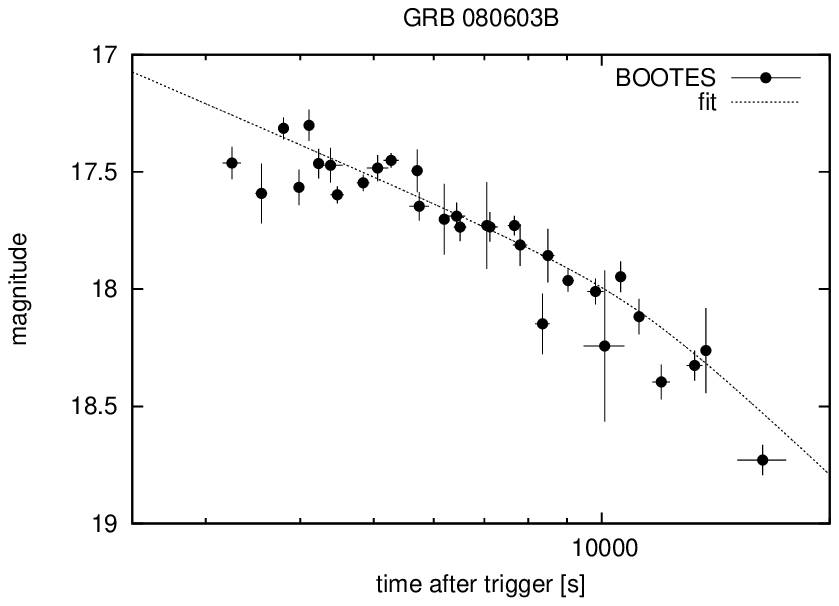}
\caption{The optical light curve of GRB\,080603B \citep{mj080603b}.
\label{bible-lc080603b}}
\end{center} 
\end{figure}

This GRB happened in Spain during sunset. We obtained first useful images
starting one hour after the trigger. An $R \simeq 17.4$ optical transient was
detected with both BOOTES-1B and BOOTES-2 see Fig.\,\ref{bible-lc080603b}.
%
BOOTES observation of this GRB is included in \citet{mj080603b}.

\paragraph{GRB\,080605}

A~long burst detected by \swift\/ \citep{rep080605}. The host was found to be
a~metal enriched star forming galaxy at redshift $z=1.64$ \citep{kruehler12}, and
exhibited the 2175\,\AA\/ extinction feature \citep{zafar12}.

\begin{figure}[b!]
\begin{center} 
\includegraphics{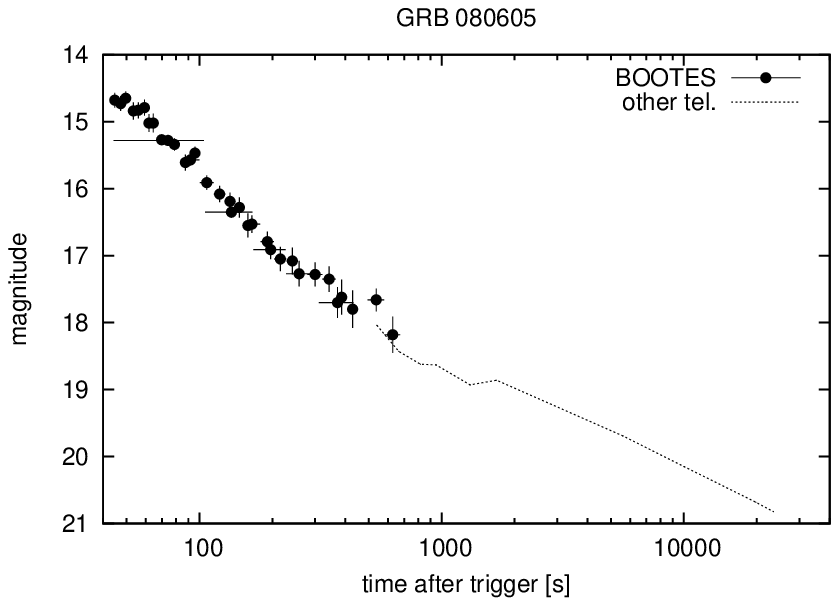}
\caption{The optical light curve of GRB\,080605 \citep{mj080605}, the dotted line is behaviour observed by \citet{crao080605} and \citet{zafar12}.
\label{bible-lc080605} }
\end{center} 
\end{figure}

GRB\,080605 was observed by both BOOTES-1B (28 photometric points) and BOOTES-2
(5 photometric points) starting 44\,s after the trigger. A~ra\-pid\-ly decaying
optical afterglow ($\alpha=1.27\pm0.04$) with R = 14.7 on the first images was
found see Fig.\,\ref{bible-lc080605}.
%
All BOOTES data are included in \citet{mj080605}.

\paragraph{GRB\,090313} 

GRB by \swift, no prompt X-rays \citep{swift090313}.  An optical afterglow
peaking at $R\sim15.6$.  Extensive optical + infrared follow-up, the first GRB
to be observed by X-Shooter.  Also detected by various observatories in radio.
Redshift $z=3.375$  \citep{aup090313,melandri090313}.

The GRB happened during daytime for BOOTES-1B and it was followed-up manually. Due
to the proximity of the Moon and limitations of then-new CCD camera driver,
many 2\,s exposures were taken to be combined later. The optical afterglow was
detected with magnitude $\sim 18.3\pm0.4$ on a~$635\times2$\,s (=21\,min)
exposure with the mid-time 11.96\,h after the GRB trigger.

\paragraph{GRB\,090813}

A~long GRB by \swift, suspected of being higher-$z$, observed also by
\sonda{Konus}-Wind and \sonda{Fermi}-GBM \citep{swift090813}. Optical
counterpart by the 1.23\,m telescope at Calar Alto with a~magnitude of
$\mathrm{I}=17.0$ \citep{goro090813}.

BOOTES-1B started observation 53\,s after the GRB, taking 10\,s unfiltered
exposures. The optical transient was weakly detected on a~combined image of
$10\times10$\,s whose exposure mean time was 630\,s after the burst. The
optical counterpart was found having $\mathrm{R}=17.9\pm0.3$. Given that the
previous and subsequent images did not show any OT detection, we might speculate
about the optical emission peaking at about this time. Also the brightness is
much weaker than might be expected from the detection by \citet{goro090813},
supporting the high redshift origin.

\begin{figure}[b!]
\begin{center} 
\includegraphics{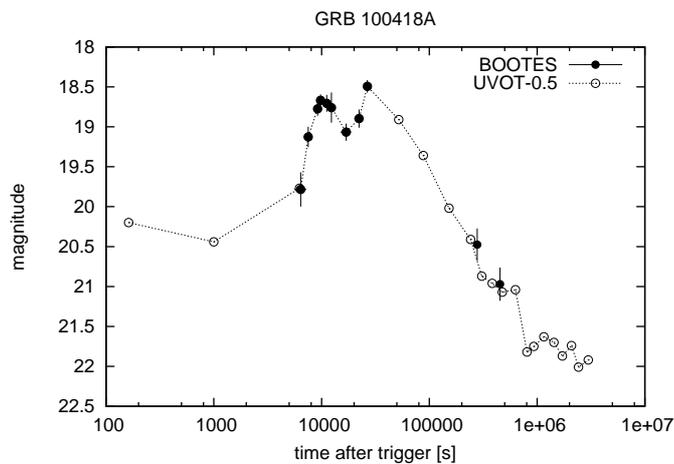}
\caption{The bizarre optical light curve of GRB\,100418A. Combination of BOOTES
and UVOT data \citep{marshall100418a}. UVOT points were shifted by an arbitrary
constant. \label{lc100418A}}
\end{center} 
\end{figure}

\paragraph{GRB\,100418A}

A~weak long burst detected by \swift\/ \citep{marshall100418a}
with a~peculiar, late-peaking optical afterglow with $z=0.6239$ \citep{aup100418a}. Also
detected in radio \citep{radio100418a}.

The first image of the GRB location was taken by BOOTES-2 at 21:50 UT (40\,min
after the GRB trigger). The rising optical afterglow was detected for the first
time on an image obtained as a~sum of 23 images, with an exposure mid-time
107\,minutes after the GRB trigger. The optical emission peaked at magnitude
R=18.7 another hour later, at an image with the mid-time 163\,min after the
trigger. A~slow decay followed, which permitted us to detect the optical
counterpart until 8 days after the GRB.\looseness=1

Because of a~mount problem, many images were lost (pointed somewhere else) and
the potential of the telescope was not fully used. Eventually, after combining
images when appropriate, 11 photometric points were obtained. A~rising part of
the optical afterglow was seen that way.

\paragraph{GRB\,100901A} 

A~long burst from \swift. Bright, slowly decaying optical afterglow discovered
by UVOT. Redshift $z=1.408$. Detected also by SMA at 345\,GHz 
\citep{rep100901a, master100901a, hartoog100901a}.

The burst happened in daytime in Spain and the position became available only
almost ten hours later after the sunset. The afterglow was still well detected
with magnitude $R \simeq 17.5$ at the beginning. BOOTES-2 had some problems with
CCD cooling, and some images were useless. The afterglow was detected also the
following night with $R = 19.35$. 

\begin{figure}[b!]
        \begin{center} 
        \includegraphics{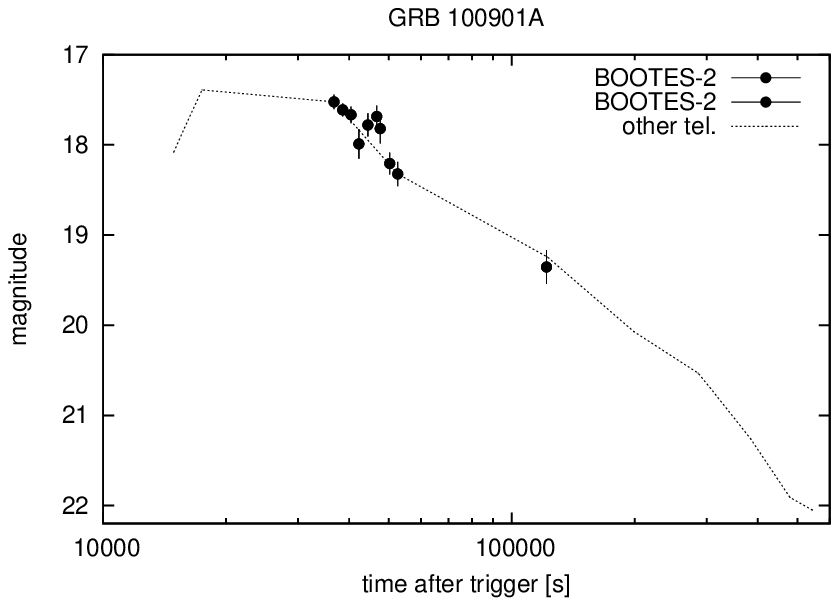}
        \caption{The optical light curve of GRB\,100901A. The dotted line representing burst behaviour is based on observations by \citet{master100901a}, \citet{kann100901a1} and \citet{crao100901a}.}
        \end{center} 
	\label{lc100901a}
\end{figure}

\begin{figure}[b!]
\begin{center} 
\includegraphics{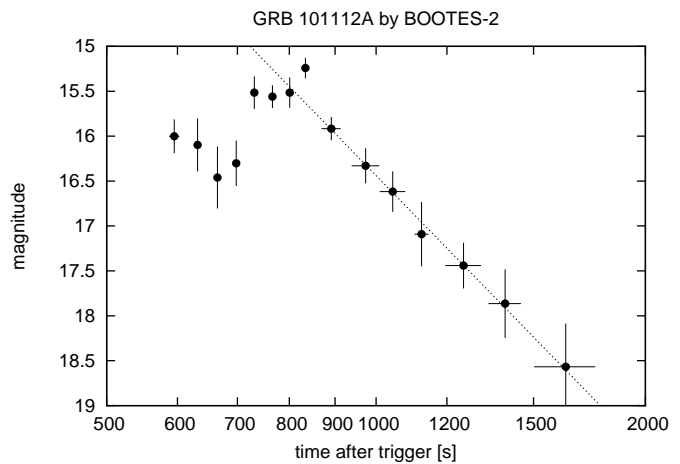}
\caption{The optical light curve of GRB\,101112A.
\label{lc101112a}}
\end{center} 
\end{figure}

\paragraph{GRB\,101112A}

An \sonda{INTEGRAL}-localized burst \citep{int101112a}, also detected by
\sonda{Fermi}-GBM \citep{gbm101112a}, \sonda{Konus}-Wind \citep{konus101112a}
and \swift-XRT \citep{xrt101112a}. Optical afterglow discovered independently
by BOOTES-2 and Liverpool Telescope \citep{lt101112a}. Detected also in radio
\citep{radio101112a}. 

BOOTES-2 reacted to the GRB101112A and started to observe 47\,s after the GRB.
A~set of 3\,s exposures was taken, but due to technical problems with the mount
a~significant amount of observing time was lost. An optical afterglow was
discovered and reported \citep{aup101112a}. The optical lightcurve exhibitted
first a~decay, then a~sudden rise to a~peak at about 800\,s after the trigger,
and finally a~surprisingly fast decay with $\alpha \simeq -4$. This behaviour
seemed more like an optical flare than a~``proper'' GRB afterglow, but there does
not seem to be contemporaneous high-energy data to make a~firm statement.

\paragraph{GRB\,110205A}

A~very long and bright burst by \swift. Detected also by \sonda{Konus}-Wind and
\sonda{Suzaku}-WAM, optical afterglow peaking at $\mathrm{R}\sim14.0$,
extensive multiwavelength follow-up, $z=2.22$ {\it ``Textbook burst''}
\citep{zheng110205a,gendre110205a}.

BOOTES-1B reacted automatically to the \swift\/ trigger. First 10\,s unfiltered
exposure was obtained 102\,s after the beginning of the GRB (with
T$_{90}=257\,s$) i.e. while the gamma-ray emission was still taking place.
After taking 18~images, the observatory triggered on a~false alarm from the
rain detector, which caused the observation to be stopped for 20~minutes.
After resuming the observation, $3\times30$\,s images were obtained and another
false alert struck over. This alert was remotely overridden by P. Kub\'{a}nek,
so that all 20~minutes were not lost. From then on, the observation continued
until sunrise. The afterglow is well detected in the images until 2.2~hours
after the GRB. 16 photometric points from combined images were eventually
published.

BOOTES-2 started observations 15\,min after the trigger, clearly detecting the
afterglow in $R$-band until 3.2\,hours after the burst. 13 photometric points
were obtained. The delay was caused by technical problems. 
BOOTES observations of this GRB are included in \citet{zheng110205a}. 

\begin{figure}[b!]
\begin{center} 
\includegraphics{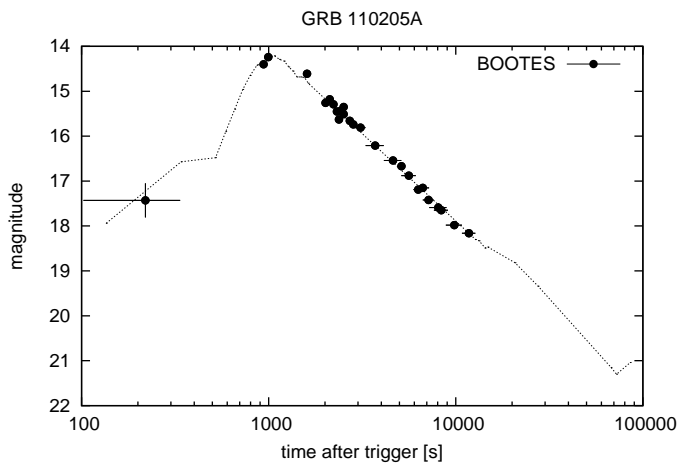}
\caption{The optical light curve of GRB\,110205A.
\label{lc110205a} }
\end{center} 
\end{figure}

\paragraph{GRB\,110213A}

A~bright burst detected by \swift, detected also by \sonda{Konus}-Wind and
\sonda{Fermi}-GBM. Optical afterglow $\mathrm{R}\sim 14.6$, extensive follow-up
\citep{swift110213a}.

BOOTES-1B started to observe 15 hours after the GRB (the position was below
horizon at the time of the trigger) and continued for an hour, eventually,
$100\times 30$\,s unfiltered images were combined, the OT brightness calibrated
against USNO-A2 is 18.3$\pm$0.3 at the exposure mid-time of 15.5\,h after the
GRB trigger.

\paragraph{GRB\,120326A} 

A~\swift-detected burst. Afterglow discovered by Tarot
\citep{2012GCN..13107...1K}.  Long-lived optical emission, redshift $z=1.78$ by
GTC. Detected also by \sonda{Fermi}-GBM and \sonda{Suzaku}-WAM \citep[][and
references therein]{swift120326a}.

At BOOTES-1B the mount failed, because of the serial port communication
failure. After a~manual recovery, 40 minutes after the GRB, images were taken
in hope for a~detection, but the counterpart with the brightness of $R\sim19.5$
was detected only at about 2$\sigma$ level. 

\paragraph{GRB\,120327A}

A~bright burst by \swift\/ with an afterglow discovered by UVOT
\citep{swift120327a}. Redshift $z=2.813$ \citep{delia120327a}. Extensive
optical follow-up.

BOOTES-1B reacted in 41\,min (similar failure as the day before: the mount
failed, because of the serial port communication failure), obtaining a~series
of 20\,s exposures. These images were combined to get 600\,s effective
exposures and permitted detection of the afterglow on six such images. The
brightness was decaying from R=17.5 to R=18.6.

All-sky camera at BOOTES-1 (CASSANDRA1) covered the event in real time and 
detected nothing down to $R\sim7.5$ \ynote{(Zanioni et al. in prep.)}.\looseness=-1

\paragraph{GRB\,121001A} A~bright and long \swift-detected GRB, originally
designated as possibly galactic \citep{swift121001a}. Afterglow discovered by
\citet{andreev121001a}. 

BOOTES-2 observed this trigger starting 32\,min after the trigger. An optical
afterglow is detected in $I$-band with $I \sim 19.7$ (Vega) for a~sum of images
between 20:49 -- 21:52\,UT \citep{bootes121001a}. 


\paragraph{GRB\,121024A} 

A~bright \swift-detected GRB with a~bright optical afterglow
\citep{swift121024a,tarot121024a}. Detected also in radio \citep{radio121024a}.
Redshift $z=2.298$ by \citet{vlt121024a}. 

BOOTES-1B observed the optical afterglow of GRB\,121024A. The observations
started 40~minutes after the GRB trigger. The sum of 20~minutes of unfiltered
images with a~mean integration time 54~minutes after the GRB shows a~weak
detection of the optical afterglow with magnitude $R=18.2 \pm 0.5$
\citep{bootes121024a}.

\paragraph{GRB\,130418A} 

A~bright and long burst with a~well detected optical afterglow somewhat
peculiarly detected after a~slew by \swift\/ \citep{swift130418a}. Observation
by \sonda{Konus}-Wind showed that the burst started already 218\,s before
\swift\/ triggered \citep{konus130418a}. Redshift $z= 1.218$ by
\citet{vlt130418a}.

BOOTES-2 obtained a~large set of unfiltered, $r'$-band and $i'$-band images
starting 1.5\,h after the trigger. The optical afterglow is well detected in
the images. The lightcurve is steadily decaying with the power-law index of
$\alpha=-0.93\pm0.06$, with the exception of the beginning, where there is
a~possible flaring with peak about 0.25\,mag brighter than the steady power-law.

\begin{figure}[b!]
\begin{center} 
\includegraphics{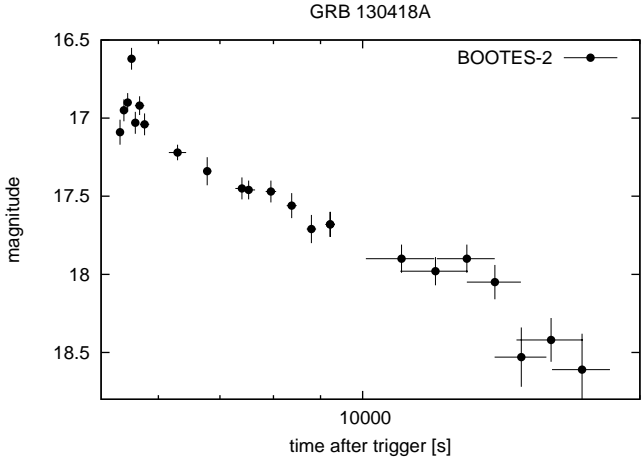}
\caption{The optical light curve of GRB\,130418A.
\label{lc130418a}}
\end{center} 
\end{figure}

\paragraph{GRB\,130505A}

A~bright and intense GRB with a~14\,mag optical afterglow detected by \swift\/
\citep{swift130505a}. Redshift $z=2.27$ reported by \citet{gemini130505a}.

BOOTES-2 obtained the first image of this GRB 11.94\,h after the trigger. A~set
of 60\,s exposures was obtained. Combining the first hour of images taken, we
clearly detect the optical afterglow, and using the calibration provided by
\citet{kann130505a}, we measure R$_C$=19.26$\pm$0.06.

\paragraph{GRB\,130606A}

A~high-redshift GRB detected by \swift\/ \citep{swift130606a}, optical
afterglow discovered by BOOTES-2, redshift $z=5.9$ by GTC \citep{ajct130606a}. 

BOOTES-2 reaction to this GRB alert was actually a~failure, the system did not
respond as well as it should and it had to be manually overridden to perform the
observations. The first image has therefore been taken as late as 13 minutes
after the trigger. These observations led to a~discovery of a~bright afterglow
not seen by \swift-UVOT, and prompted spectroscopic observations by 10.4m GTC,
which show redshift of this event to be $z=5.9135$. Overall, 14 photometric
points in $i'$-band and 7 in $z'$-band were obtained \citep{ajct130606a}.

\begin{figure}[b!]
\begin{center} 
\includegraphics{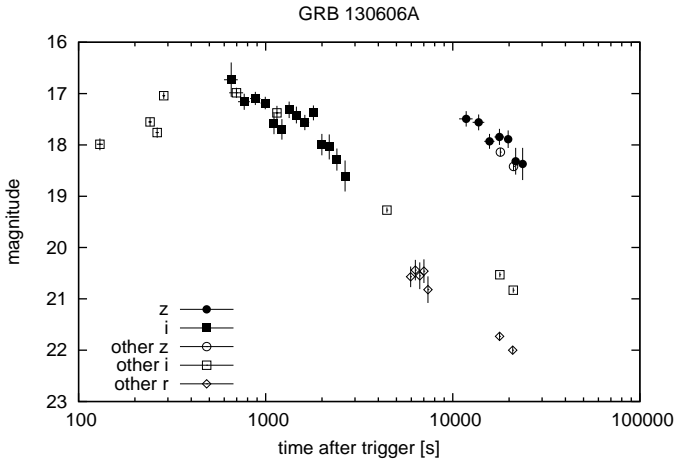}
\caption{The optical light curve of GRB\,130606A. $i'$-band points were shifted
2.4\,mag up to match with  the $z'$-band points. \label{lc130606a}}
\end{center} 
\end{figure}

\begin{deluxetable}{lcccc}
\tablewidth{0pt}
\tablecaption{BOOTES-1B GRBs in a~table
\label{table-bible-b1}}
\tabletypesize{\small}
\tablehead{ \colhead{GRB} & \colhead{$\Delta T$} & \colhead{no. pts} &
\colhead{result} & \colhead{ref.} }
\startdata
030913  & 2\,h  &   & $V>17.5$, $C>12$   &        \\
050215B & 22\,m &   & $V>16.5$, $I>15.0$ &        \\
050505  & 47\,m &   & $V>19$             &        \\
050509A & 64\,m &   & $V>14.9$           &        \\
050509B & 62\,s &   & $V>11.5$           &        \\
050525A & 12\,m$^\dagger$ & 1 & $16.5\pm0.4$       & [1] \\
050528  & 71\,s &   & $V>13.8$, $I>13.0$ &        \\
050824  & 10\,m & 4 & $R=18.2\pm0.3$     &  [2]     \\
050904  & 2\,m  &   & $R>18.2$           &  [3]  \\
050922C & 4\,m  & 3 & $R=14.6\pm0.4$     &        \\
051109A & 55\,s & 6 & $R=15.7\pm0.4$     &        \\
051211B & 42\,s &   & $R>$               &        \\
051221B & 4\,m  &   & $R>$               &        \\
060421  & 61\,s &   & $R>14$             &        \\
061110B & 11\,m &   & $R>18$             &        \\
071101  & 55\,s &   & $C>17.0$           &        \\
071109  & 59\,s &   & $C>13.0$           &        \\
080330  &  6\,m & 6 & $C=16.5\pm 0.2$    &        \\
080413A & 61\,s & 61& $C\simeq13.3$      &        \\
080430  & 34\,s & * & $C\simeq15.5$      &        \\
080603B &  1\,h & 11& $C\simeq17.4$      &   [4]    \\
080605  & 44\,s & 28& $C\simeq14.7$      &   [5]    \\
081003B & 41\,s &   & $C>17.6$           &        \\
090313  & 12\,h & 1 & $C\simeq18.3$      &        \\
090519  & 99\,s &   & $C>17.6$           &        \\
090813  & 53\,s & 1 & $C\simeq17.9$      &        \\
090814A &  3\,m$^\dagger$ &   & $C>15.8$           & \\
090814B & 53\,s$^\dagger$ &   & $C>17.5$           & \\
090817  & 24\,m &   & $C>16.7$           &        \\
100906A & 106\,s&   & $C>16.5$           &        \\
110205A & 102\,s&16 & $C\sim14$          &  [6]    \\
110212A &  50\,s&   & $C>13.0$           &        \\
110213A & 15\,h & 1 & $C=18.3\pm0.2$     &        \\
110411A & 24\,s &   & $C>17.8$           &        \\
111016A &1.25\,h&   & $C>17.8$           &        \\
120326A & 40\,m & 1 & $C\sim19.5$        &        \\
120327A & 41\,m$^\dagger$ & 6 & $C=17.5$           & \\
120328A & 7.5\,m&   & $C>16$             &        \\
120521C &11.7\,m&   & $C>20.5$           &        \\
120711B &107\,s &   & $C>18.2$           &        \\
120729A &10\,h  &   & $C>19.0$           &        \\
121017A & 79\,s &   & $C>19.0$           &        \\
121024A & 40\,m & 1 & $C=18.2\pm0.5$     &        \\
121209A & 42\,s$^\dagger$ &   & $C>16.5$           & \\
130122A & 28\,m &   & $C>18.4$           &        \\
\enddata
\tablecomments{ 1. \citet{resmi050525}, 2. \citet{js050824},  3.
\citet{nature050904}, 4. \citet{mj080603b}, 5. \citet{mj080605}, 6.
\citet{zheng110205a}, $^\dagger$ marks alerts covered in real time by
wide-field camera CASSANDRA-1.}
\end{deluxetable}

\begin{deluxetable}{lcccc}
\tablewidth{0pt}
\tablecaption{BOOTES-2 GRBs in a~table
\label{table-bible-b2}
}
\tabletypesize{\small}
\tablehead{ \colhead{GRB} & \colhead{$\Delta T$} & \colhead{no. pts} & \colhead{result} & \colhead{ref.} }
\startdata

080603B &         & 20& $R\simeq17.4 $   &   [1]      \\
080605  &         & 5 & $R\simeq14.7 $   &   [2]  \\
090817  & 145\,s  &   & $R>18.3$         &         \\
090904A &  86\,s  &   & $R>16.1$         &         \\
091202  & 5.5\,h  &   & $>18.3$          &         \\
100219A & 6.3\,h  &   & $C>18.3$         &         \\
100418A & 1.8\,h  &11 & $C=19.3$         &         \\
100522A & 625\,s  &   & $C>15.5$         &         \\
100526A & 4\,h    &   & $r'>14$          &         \\
100614A & 6.9\,m  &   & $C>18$           &         \\
100901A & 10\,h  &10 & $C=17.52\pm0.08$ &         \\
100915A & 106\,s  &   & $C>16.5$         &         \\
101020A & 5.1\,h  &   & $r'>18.0$        &         \\
101112A & 595\,s  &15 & $C=15.5$         &         \\
110106B & 10.3\,m &   & $C>16.5$         &         \\
110205A & 15\,m   &13 & $R\sim14$        & [3] \\
110212A & 32\,m   &   & $R>16.5$         &         \\
110223A & 228\,s  &   & $R>17.6$         &         \\
120729A &13.25\,h &   & $R>19.4$         &         \\
120805A &   25\,m &   & $R>18.5$         &         \\
120816A &   66\,m &   & $R>18$           &         \\
121001A &   32\,m &   & $I>19.7$         &         \\
121017A &   3\,m  &   & $C>18.5$, $i'>19.5$ &      \\
130418A & 1.5\,h  & 21& $C=16.8\pm0.06$  &         \\
130505A &11.94\,h & 1 & $R_C = 19.26 \pm 0.06$ & \\
130606A &  13\,m  &21 & $i'=16.7\pm0.3 $ &         \\
130608A & 2.3\,h  &   & $C>18.8$         &         \\ 
130612A & 4.8\,m  &   & $C>18.6$         &         \\ 
130806A & 40\,s   &   & $C>18.3$         &         \\ 
131202A & 4.25\,h &   & $i'>19.7$        &         \\

\enddata
\tablecomments{
1. \cite{mj080603b}, 2. \citet{mj080605},  3. \citet{zheng110205a}}
\end{deluxetable}

\section{Summary
\label{bible-concl}}

Eleven years of BOOTES-1B and BOOTES-2 GRB follow-up history are summarised in
the textual and tabular form. 
Each GRB is given a~short introductory paragraph as a~reminder of the basic
optical properties of the event.  Although we do not discuss the properties in
other wavelengths, we try to include a~comprehensive reference of literature
relevant to each burst.  One by one, we show all the successful follow-ups that
these telescopes have performed during the first ten years of the \swift\/ era,
and the transition of the BOOTES network to the RTS-2 \citep{rts2} observatory
control system, first installed at BOOTES-2 in 2003, and made definitive during
the summer of 2004.

The BOOTES telescopes, in spite of their moderate apertures ($\simlt 60$\,cm)
have proven to detect a significant number of afterglows -- together over 20,
contributing to the understanding of the early GRB phase. 

{\small \paragraph{Acknowledgements}

This paper is dedicated to the memory of our colleagues Dolores Perez Ramirez y
Javier Gorosabel, both of whom passed away while this paper was in preparation.
They were far too young, and are greatly missed.

We appreciate the auspices of INTA, IHSM-UMA/CSIC and UMA as well as the
financial support by the Junta de Andaluc\'{\i}a and the Spanish Ministry of
Economy and Competitiveness through the research projects P07-TIC-03094,
P12-TIC2839, AYA 2009-14000-C03-01, AYA 2010-39727-C03-01 and AYA-2015-71718-R. 
MJ was supported by the postdoctoral fellowship of the Czech Academy of
Sciences. 
This study was carried out in the framework of the Unidad Asociada IAA-CSIC at
the group of planetary science of ETSI-UPV/EHU.  
This work was supported by the Ikerbasque Foundation for Science.
The Czech CVUT FEL team acknowledges the support by GA CR grants 
13-33324S 


\bibliography{tes}
\bibliographystyle{aa}

\begin{deluxetable}{lcccl}
\tablewidth{0pt}
\tabletypesize{\normalsize}
\tablecaption{GRB\,050525A: Observing log of BOOTES-1B
\label{bible-phot050525a}}
\tablehead{ \colhead{$\Delta T[h]$} & \colhead{exp[s]} & \colhead{mag} & \colhead{dmag} & \colhead{filter} }
\startdata
0.195 & $39 \times 10$\,s & 16.51 & 0.39 & R \\
\enddata
\tablecomments{Published by \citet{resmi050525}} \\
\end{deluxetable}

\begin{deluxetable}{lcccc}
\tablewidth{0pt}
\tabletypesize{\normalsize}
\tablecaption{GRB\,050824: Observing log of BOOTES-1B
\label{bible-phot050824}}
\tablehead{ \colhead{$\Delta T[h]$} & \colhead{exp[s]} & \colhead{mag} & \colhead{dmag} & \colhead{filter} }
\startdata
0.1763 & & 18.22 & 0.35 & R \\ 
0.3462 & & 19.11 & 0.32 & R \\
1.0249 & & 19.67 & 0.33 & R \\
2.9091 & & 19.33 & 0.20 & R \\
\enddata
\tablecomments{Published by \citet{js050824}} \\
\end{deluxetable}

\begin{deluxetable}{lcccl}
\tablewidth{0pt}
\tabletypesize{\normalsize}
\tablecaption{GRB\,050922C: Observing log of BOOTES-1B
\label{bible-phot050922C}}
\tablehead{ \colhead{$\Delta T[h]$} & \colhead{exp[s]} & \colhead{mag} & \colhead{dmag} & \colhead{filter} }
\startdata
0.0694 & 40 & 14.58 & 0.35 & R \\
0.3752 & 900 & 17.01 & 0.39 & R \\
0.6193 & 900 & 18.53 & 0.59 & R \\
\enddata
\end{deluxetable}
\begin{deluxetable}{lcccc}
\tablewidth{0pt}
\tabletypesize{\normalsize}
\tablecaption{GRB\,051109A: Observing log of BOOTES-1B
\label{bible-phot051109A}}
\tablehead{ \colhead{$\Delta T[s]$} & \colhead{exp[s]} & \colhead{mag} & \colhead{dmag} & \colhead{filter} }
\startdata
59.7 & 10 & 15.67 & 0.35 & R \\
122.2 & 74 & 16.02 & 0.19 & R \\
257.9 & 41 & 16.65 & 0.41 & R \\
756.6 & 205 & 17.18 & 0.22 & R \\
1021.5 & 313 & 17.68 & 0.26 & R \\
508.4 & 908 & 16.98 & 0.54 & I \\ 
\enddata
\end{deluxetable}

\begin{deluxetable}{lcccl}
\tablewidth{0pt}
\tabletypesize{\normalsize}
\tablecaption{GRB\,080330: Observing log of BOOTES-1B
\label{bible-phot080330}}
\tablehead{ \colhead{$\Delta T[h]$} & \colhead{exp[s]} & \colhead{mag} & \colhead{dmag} & \colhead{filter} }
\startdata
0.1061 & 7   & 16.52 & 0.23 & clear \\
0.3752 & 210 & 16.61 & 0.13 & clear \\
0.6193 & 588 & 17.16 & 0.14 & clear \\
0.8547 & 825 & 17.45 & 0.15 & clear \\
1.0915 & 862 & 17.29 & 0.13 & clear \\
1.3384 & 905 & 17.42 & 0.16 & clear \\
\enddata
\end{deluxetable}

\begin{deluxetable}{lcccl}
\tablewidth{0pt}
\tabletypesize{\normalsize}
\tablecaption{GRB\,090813: Observing log of BOOTES-1B
\label{bible-phot090813}}
\tablehead{ \colhead{$\Delta T[h]$} & \colhead{exp[s]} & \colhead{mag} & \colhead{dmag} & \colhead{filter} }
\startdata
0.175 & $10\times10$ & 17.9 & 0.3 & clear \\
\enddata
\end{deluxetable}

\begin{deluxetable}{lcccl}
\tablewidth{0pt}
\tabletypesize{\normalsize}
\tablecaption{GRB\,100418A: Observing log of BOOTES-2
\label{bible-phot100418A}}
\tablehead{ \colhead{$\Delta T[h]$} & \colhead{exp[s]} & \colhead{mag} & \colhead{dmag} & \colhead{filter} }
\startdata
1.78  & 1638  & 19.785 & 0.215 & clear \\
2.09  & 597   & 19.127 & 0.127 & clear \\
2.55  & 534   & 18.774 & 0.087 & clear \\
2.72  & 656   & 18.668 & 0.073 & clear \\
3.10  & 239   & 18.706 & 0.106 & clear \\
3.43  & 238   & 18.759 & 0.189 & clear \\
4.70  & 3908  & 19.067 & 0.108 & clear \\
6.19  & 4328  & 18.897 & 0.115 & clear \\
7.39  & 551   & 18.493 & 0.078 & clear \\
77.3  & 14830 & 20.475 & 0.202 & clear \\
125.6 & 12482 & 20.970 & 0.208 & clear \\
\enddata
\end{deluxetable}

\begin{deluxetable}{lcccl}
\tablewidth{0pt}
\tabletypesize{\normalsize}
\tablecaption{GRB\,100901A: Observing log of BOOTES-2
\label{bible-phot100901A}}
\tablehead{
\colhead{$\Delta T[h]$} & \colhead{exp[s]} &  \colhead{mag} & \colhead{dmag} & \colhead{filter} }
\startdata
10.202 &  268 & 17.52 & 0.08 & R \\
10.719 &  415 & 17.61 & 0.07 & R \\
11.230 &  354 & 17.67 & 0.09 & R \\
11.734 &  238 & 17.99 & 0.16 & R \\
12.346 &  730 & 17.78 & 0.13 & R \\
12.980 &  759 & 17.68 & 0.12 & R \\
13.239 &  759 & 17.82 & 0.16 & R \\
13.971 &  997 & 18.21 & 0.12 & R \\
14.611 & 1101 & 18.32 & 0.14 & R \\
33.791 & 4012 & 19.35 & 0.19 & R \\
\enddata
\end{deluxetable}

\begin{deluxetable}{lcccc}
\tablewidth{0pt}
\tabletypesize{\small}
\tablecaption{GRB\,101112A: Observing log of BOOTES-2
\label{bible-phot101112A}}
\tablehead{ \colhead{$\Delta T[s]$} & \colhead{exp[s]} &  \colhead{mag} &
\colhead{dmag} & \colhead{filter} }
\startdata
595.0  & 16  & 16.00 & 0.19 & r' \\
631.8  & 8   & 16.10 & 0.29 & r' \\
664.9  & 7   & 16.46 & 0.34 & r' \\
697.8  & 7   & 16.30 & 0.25 & r' \\
731.0  & 7   & 15.52 & 0.18 & r' \\
766.1  & 11  & 15.56 & 0.13 & r' \\
800.9  & 7   & 15.52 & 0.17 & r' \\
833.8  & 7   & 15.24 & 0.12 & r' \\
891.2  & 44  & 15.92 & 0.13 & r' \\
973.7  & 69  & 16.33 & 0.20 & r' \\
1044.0 & 69  & 16.62 & 0.23 & r' \\
1124.2 & 41  & 17.09 & 0.36 & r' \\
1252.7 & 115 & 17.44 & 0.25 & r' \\
1393.8 & 116 & 17.86 & 0.38 & r' \\
1629.5 & 255 & 18.57 & 0.48 & r' \\
\enddata
\end{deluxetable}

\begin{deluxetable}{lcccl}
\tablewidth{0pt}
\tabletypesize{\normalsize}
\tablecaption{GRB\,110213A: Observing log of BOOTES-1B
\label{bible-phot110213A}}
\tablehead{ \colhead{$\Delta T[h]$} & \colhead{exp[s]} &  \colhead{mag} &
\colhead{dmag} & \colhead{filter} }
\startdata
15.5 & $100\times30$ & 18.29 & 0.30 & clear \\
\enddata
\end{deluxetable}

\begin{deluxetable}{lcccl}
\tablewidth{0pt}
\tabletypesize{\normalsize}
\tablecaption{GRB\,120327A: Observing log of BOOTES-1B
\label{bible-phot120327A}}
\tablehead{
\colhead{$\Delta T[h]$} & \colhead{exp[s]} &  \colhead{mag} & \colhead{dmag} & \colhead{filter} }
\startdata
0.955 & 654 & 17.50 & 0.12 & clear \\
1.140 & 674 & 17.65 & 0.12 & clear \\
1.337 & 748 & 17.82 & 0.13 & clear \\
1.533 & 660 & 18.24 & 0.21 & clear \\
1.718 & 673 & 18.17 & 0.21 & clear \\
1.905 & 656 & 18.59 & 0.29 & clear \\
\enddata
\end{deluxetable}

\begin{deluxetable}{lcccl}
\tablewidth{0pt}
\tabletypesize{\normalsize}
\tablecaption{GRB\,121024A: Observing log of BOOTES-1B
\label{bible-phot121024A}}
\tablehead{ \colhead{$\Delta T[h]$} & \colhead{exp[s]} &  \colhead{mag} &
\colhead{dmag} & \colhead{filter} }
\startdata
0.900 & 1200 & 18.2 & 0.5 & clear \\
\enddata
\end{deluxetable}

\begin{deluxetable}{lcccc}
\tablewidth{0pt}
\tabletypesize{\normalsize}
\tablecaption{GRB\,130418A: Observing log of BOOTES-1B and BOOTES-2
\label{bible-phot130418A}}
\tablehead{ \colhead{$\Delta T[h]$} & \colhead{exp[s]} &  \colhead{mag} &
\colhead{dmag} & \colhead{filter} }
\startdata
1.514 & $3 \times 15$\,s & 17.09 & 0.08 & clear\\
1.529 & $3 \times 15$\,s & 16.95 & 0.07 & clear\\
1.544 & $3 \times 15$\,s & 16.90 & 0.06 & clear\\
1.558 & $3 \times 15$\,s & 16.62 & 0.07 & clear\\
1.573 & $3 \times 15$\,s & 17.03 & 0.07 & clear\\
1.590 & $4 \times 15$\,s & 16.92 & 0.06 & clear\\
1.610 & $4 \times 15$\,s & 17.04 & 0.07 & clear\\
1.749 & $7 \times 15$\,s & 17.22 & 0.05 & clear\\
1.865 &            60\,s & 16.92 & 0.18 & $r'$ \\
1.884 & $4 \times 15$\,s & 17.34 & 0.09 & clear\\
2.054 & $7 \times 15$\,s & 17.45 & 0.07 & clear\\
2.089 & $7 \times 15$\,s & 17.46 & 0.06 & clear\\
2.209 & $6 \times 15$\,s & 17.47 & 0.07 & clear\\
2.326 & $6 \times 15$\,s & 17.56 & 0.08 & clear\\
2.444 & $6 \times 15$\,s & 17.71 & 0.09 & clear\\
2.562 & $6 \times 15$\,s & 17.68 & 0.08 & clear\\
2.798 & $22 \times 60$\,s & 17.40 & 0.04 & $i'$ \\
3.061 & $15 \times 60$\,s & 17.90 & 0.09 & $r'$ \\
3.333 & $15 \times 60$\,s & 17.98 & 0.09 & $r'$ \\
3.604 & $15 \times 60$\,s & 17.90 & 0.09 & $r'$ \\
3.866 & $15 \times 60$\,s & 18.05 & 0.11 & $r'$ \\
4.130 & $15 \times 60$\,s & 18.53 & 0.19 & $r'$ \\
4.449 & $20 \times 60$\,s & 18.42 & 0.14 & $r'$ \\
4.808 & $20 \times 60$\,s & 18.61 & 0.23 & $r'$ \\
\enddata
\end{deluxetable}

\begin{deluxetable}{lcccl}
\tablewidth{0pt}
\tabletypesize{\normalsize}
\tablecaption{GRB\,130505A: Observing log of BOOTES-2
\label{bible-phot130505A}}
\tablehead{ \colhead{$\Delta T[h]$} & \colhead{exp[s]} &  \colhead{mag} &
\colhead{dmag} & \colhead{filter} }
\startdata
12.488 & $51 \times 60$\,s & 19.26 & 0.06 & clear \\
\enddata
\end{deluxetable}

\end{document}